\documentclass{article}

\PassOptionsToPackage{numbers, sort&compress}{natbib}

\usepackage[final]{nips_2018}
\usepackage[utf8]{inputenc} 
\usepackage[T1]{fontenc}    
\usepackage{hyperref}       
\usepackage{url}            
\usepackage{booktabs}       
\usepackage{amsfonts}       
\usepackage{amsmath}
\usepackage{nicefrac}       
\usepackage{microtype}      
\usepackage{graphicx}       
\usepackage[bottom]{footmisc} 
\usepackage{tablefootnote}
\usepackage{multirow}
\usepackage{dcolumn}
\usepackage{tabularx}
\usepackage{caption}
\usepackage{booktabs}

\usepackage[dvipsnames]{xcolor}

\usepackage[normalem]{ulem}

\newcommand{\norm}[1]{\left\lVert#1\right\rVert}

\title{Analysis of Atomistic Representations Using Weighted Skip-Connections}

\author{
  Kim A.~Nicoli\\
  Machine Learning Group\\
  Technische Universit\"at Berlin\\
  10587 Berlin, Germany\\
  \texttt{kim.a.nicoli@tu-berlin.de}
  \and
   \textbf{Pan Kessel}\\
  	Machine Learning Group\\
  	Technische Universit\"at Berlin\\
  	10587 Berlin, Germany\\
  	\vspace{1.0cm}
  \and
    \textbf{Michael Gastegger}\\
 Machine Learning Group\\
 Technische Universit\"at Berlin\\
 10587 Berlin, Germany\\
 \and
\textbf{Kristof T. Sch\"utt}\\
Machine Learning Group\\
Technische Universit\"at Berlin\\
10587 Berlin, Germany\\
\texttt{kristof.schuett@tu-berlin.de}}
\begin{document}

\maketitle
\begin{abstract}
In this work, we extend the SchNet architecture by using weighted skip connections to assemble the final representation. This enables us to study the relative importance of each interaction block for property prediction. We demonstrate on both the QM9 and MD17 dataset that their relative weighting depends strongly on the chemical composition and configurational degrees of freedom of the molecules which opens the path towards a more detailed understanding of machine learning models for molecules.
\end{abstract}

\section{Introduction}
In recent years, machine learning has found wide-spread and successful application in quantum chemistry, condensed-matter physics and materials science, e.g. for potential energy surface and accelerated molecular dynamics simulations~\citep{behler2007generalized,gastegger2017machine,behler_first_2017,glielmo2017accurate,chmiela_machine_2017,schnet_JCP,chmiela2018towards} as well as predictions of properties across chemical compound space~\citep{rupp2012fast,de2016comparing,faber2017prediction,podryabinkin2017active,bartok2017machine,dragoni2018achieving,faber2018alchemical,pronobis2018capturing,pronobis2018many}. 
Recent deep learning architectures~\cite{DTNN,Schnet_NIPS,lubbers2018hierarchical,gilmer2017neural,jorgensen2018neural} yield accurate predictions of chemical properties while learning atomistic representations directly from atom types and positions.
Beyond the accuracy of those networks, there has been increasing research regarding the interpretation of their  predictions~\cite{baehrens2010explain,simonyan2013deep,zeiler2014visualizing,LRP2,kindermans2018learning} as well as extracting quantum-chemical insights from learned atomistic representations~\cite{DTNN,schnet_JCP,Book_chapter}.
SchNet~\cite{schnet_JCP} and DTNN~\cite{DTNN} use a sequence of interaction blocks to model quantum interactions.
While in a single interaction block only pairwise interactions are considered, their repeated application infuses increasingly complex environmental information into the atom-wise representations.
Intuitively, the first layers can only model pair-wise, local interactions, whereas deeper layers are able to capture more complex, longer ranging interactions.

In this paper, we aim to quantify how much each interaction block contributes to the final representation depending on the underlying reference data as well as how the importance of interaction blocks develops during training.
We achieve this by introducing a modified SchNet architecture -- sc-SchNet -- which uses weighted skip-connections to assemble the final representation (see Fig.~\ref{fig:figure_01}).
Skip-connections have been shown to smoothen the loss landscape and enable the training of extremely deep networks~\cite{li2018visualizing, he_deep_2015, huang_densely_2016}. 
While SchNet already contains ResNet-style skip-connections by modeling the interactions as additive corrections, here we add skip-connections by assembling the final atom-wise representations as a linear combination of the intermediate representations obtained by each interaction block.
This allows us to explore how SchNet obtains molecular representations from these interactions and how the underlying data influence the training process as well as the final representation.

The remainder of this work is structured as follows: first we give an overview of the original SchNet architecture~\cite{schnet_JCP, Schnet_NIPS} and outline the modifications that lead to sc-SchNet. Then, we present results for energy predictions across chemical compound space using the QM9 dataset and also for prediction of potential energy surfaces and force fields using the MD17 dataset. Finally, we analyze the weighted skip-connections to obtain insights about how the structure of a molecule as well as the composition of a dataset influences the learned representations.

\section{Method}
\begin{figure}[t]
  \centering
  \includegraphics[width=0.7\textwidth]{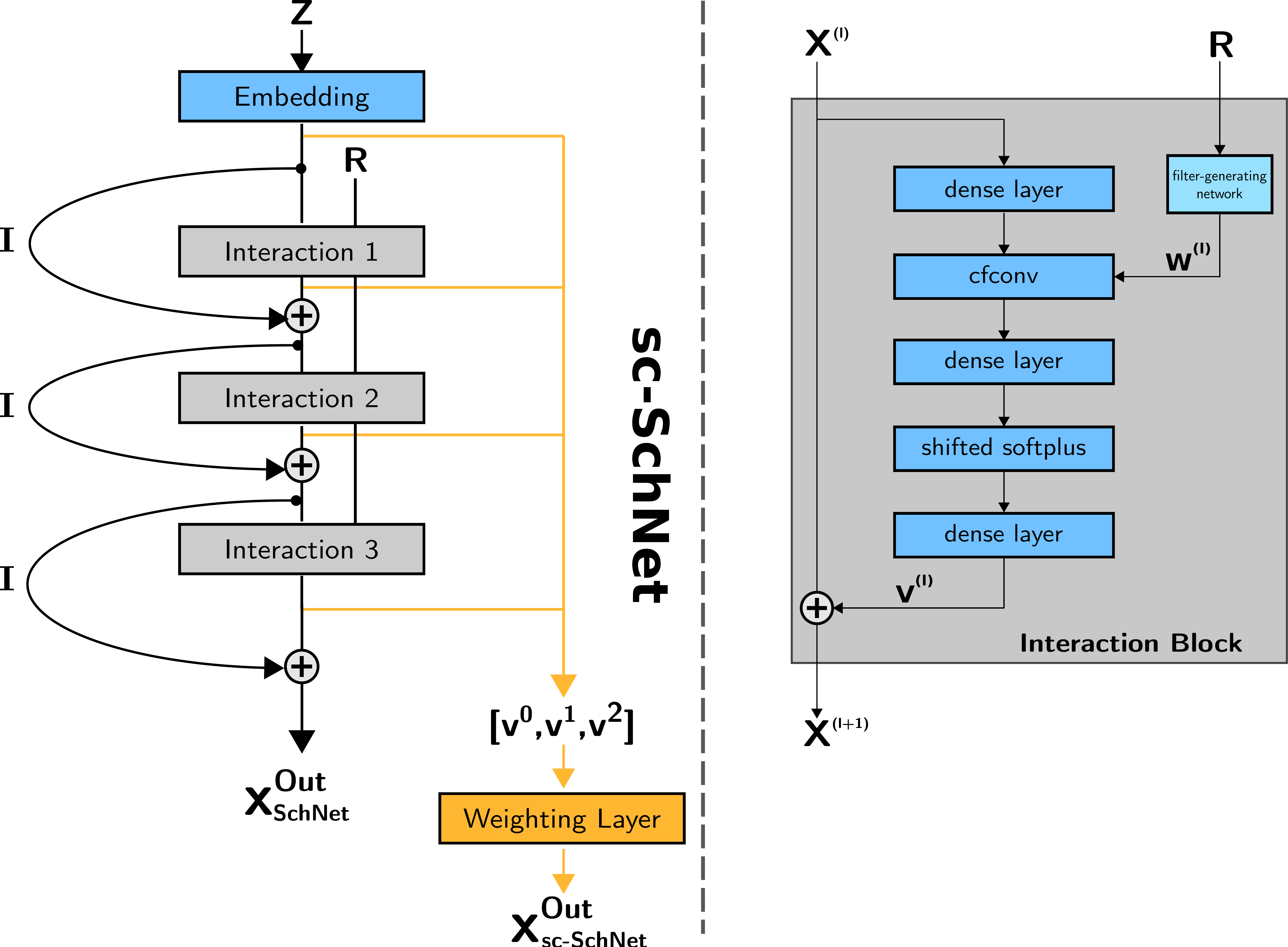}
  \caption[SchNet vs sc-SchNet]{\small An overview of the differences between standard SchNet and sc-SchNet is shown on the left. We use black and orange arrows for the former and latter architecture, respectively. The structure of an interaction block is shown on the right. We refer to \cite{schnet_JCP} for further details.}
  \label{fig:figure_01}
\end{figure}
SchNet~\cite{schnet_JCP} is a deep neural network for atomistic systems following the DTNN framework~\cite{DTNN}, 
i.e. atom-wise representations are constructed by starting from element-specific embeddings $\mathbf{x}_i^{(0)} = \mathbf{a}_{Z_i}$, for atom $i$ with nuclear charge $Z_i$, followed by repeated, pair-wise interaction corrections
\begin{equation}\label{eq:update}
\mathbf{x}_i^{(l)} = \mathbf{x}_i^{(l-1)} + \mathbf{v}_i^{(l)}.
\end{equation}
Fig.~\ref{fig:figure_01} shows an overview of the SchNet using $T=3$ interaction blocks on the left as well as the architecture of the interaction block $\mathbf{v}_i^{(l)}$ on the right.
Because the corrections $\mathbf{v}_i^{(l)}$ depend on all previous atom-wise representations $\mathbf{x}_j^{(l-1)}$ of neighbors $j$ within a given cutoff radius, more and more complex spatial information is incorporated in the atom-wise representations $\mathbf{x}_i^{(l)}$ .
SchNet models quantum interactions by continuous-filter convolutions~\cite{Schnet_NIPS} in order to enable smooth predictions of potential energy surfaces.
After a given number $L$ of interaction corrections, the energy of a molecule is predicted by
\begin{equation}
E = \sum_{i=1}^n g(\mathbf{x}_i^{\mathrm{out}}) ,
\end{equation}
where a fully-connected neural network $g$ predicts latent atom-wise energy contributions from the final representations $\mathbf{x}_i^{\mathrm{out}}$.
For further details, we refer to the original papers~\cite{Schnet_NIPS,schnet_JCP} as well as the reference implementation~\cite{Schnetpack}.

In this work, we extend this architecture by weighted skip-connections. 
More concretely, we do not only pass the atom features through the interactions blocks but also construct the final representation as a weighted sum of intermediate representations.
This allows the model to access the evolution of the features through the entire forward pass.
In order to distinguish our extended architecture from standard SchNet, we will refer to it as \textit{sc-SchNet} in the following.
Fig.~\ref{fig:figure_01} (left) visualizes the modifications applied to SchNet to arrive at sc-SchNet in orange.

First, we unroll the interaction corrections of the standard SchNet, i.e. Eq.~\ref{eq:update}, as
\begin{equation}\label{eq:sum}
\mathbf{x}^{\mathrm{out}}_i=\sum_{l=0}^{L} \mathbf{v}^{(l)}_i \, ,
\end{equation}
defining $\mathbf{v}_i^{(0)}=\mathbf{x}_i^{(0)}$ for compactness.
In sc-SchNet, we instead use a weighted sum over interaction corrections
\begin{equation}\label{eq:weightsum}
\mathbf{x}_i^{\mathrm{out}}=\sum_{l=0}^{L} {w}^{(l)}\cdot \mathbf{v}_i^{(l)} \, ,
\end{equation}
where the $w^{l}$ are \emph{contribution weights} of the different interaction stages.
Contribution weights are trainable parameters of the network. They are initialized as uniformly distributed, and normalized to one, before being updated during the learning process.

Note that the interaction corrections $\mathbf{v}_i^{(l)}$ still only depend on the previous corrections through Eq.~\ref{eq:update}.
Therefore, sc-SchNet effectively decouples the evolution of the features during the interaction corrections from the composition of the final representation.
While this may help during training~\cite{li2018visualizing}, it also allows us to obtain interpretable contributions of different interaction blocks.

This approach however has the disadvantage that the contribution weights depend on the magnitude of the interaction correction $\mathbf{v}_i$. In order to avoid this, we rescale them with respect to this magnitude by replacing the aggregation rule \eqref{eq:weightsum} by
\begin{equation}\label{eq:weightsum_2}
\mathbf{x}_i^{\mathrm{out}}=\sum_{l=0}^{L} \norm{\mathbf{v}}_{\textrm{Avg}} {w}^{(l)}\cdot \frac{\mathbf{v}_i^{(l)}}{\norm{\mathbf{v}}_{\textrm{Avg}}}=\sum_{l=0}^{L} \hat{w}^{(l)}\cdot \frac{\mathbf{v}_i^{(l)}}{\norm{\mathbf{v}}_{\textrm{Avg}}} \,,
\end{equation}
where we defined the quantity $\hat{w}^{(l)}=\norm{\mathbf{v}}_{\textrm{Avg}}\cdot{w}^{(l)}$ with $\norm{\mathbf{v}}_{\textrm{Avg}}$ denoting the average norm in the current minibatch.
In the following sections, we will only consider the normalized contribution weights $\hat{w}^{(l)}$.

\section{Results}\label{sec:result}
\begin{table}[tb]
	\caption{Summary of performance on the test set. By $N$, we denote the size of the combined train and validation set. Best results in \textbf{bold}. \label{tab:results}}
	\centering
	\begin{tabular}{llllcc} 
		\toprule
		Dataset & Property & Unit & Model & MAE & RMSE \\
		\midrule
		\multirow{4}{*}{Aspirin (N=1k)} &  \multirow{2}{*}{energy} & \multirow{2}{*}{kcal mol$^{-1}$} & SchNet & 0.40 $\pm$ 0.01 & 0.57 $\pm$ 0.01  \\ 
		& & &  sc-SchNet & \textbf{0.39} $\pm$ 0.01  & \textbf{0.54} $\pm$ 0.01  \\
		&  \multirow{2}{*}{forces} & \multirow{2}{*}{kcal mol$^{-1}$ {\AA}$^{-1}$} & SchNet & \textbf{0.94} $\pm$ 0.01 & \textbf{ 1.38} $\pm$ 0.02 \\ 
		& & &  sc-SchNet & 0.95 $\pm$ 0.01  & 1.47 $\pm$ 0.02 \\ 
		\midrule
		
		\multirow{4}{*}{Salicylic Acid (N=1k)} &  \multirow{2}{*}{energy} & \multirow{2}{*}{kcal mol$^{-1}$} & SchNet & 0.24 $\pm$ 0.01 & 0.31 $\pm$ 0.01  \\ 
		& & &  sc-SchNet & \textbf{0.23} $\pm$ 0.01  & \textbf{0.29} $\pm$ 0.01  \\
		&  \multirow{2}{*}{forces} & \multirow{2}{*}{kcal mol$^{-1}$ {\AA}$^{-1}$} & SchNet & \textbf{0.63} $\pm$ 0.01 & \textbf{ 0.94} $\pm$ 0.01 \\ 
		& & &  sc-SchNet & 0.63 $\pm$ 0.01  & 0.97 $\pm$ 0.02 \\ 
		\midrule
		
		\multirow{4}{*}{Benzene (N=1k)} &  \multirow{2}{*}{energy} & \multirow{2}{*}{kcal mol$^{-1}$} & SchNet & 0.10 $\pm$ 0.00 & 0.12 $\pm$ 0.01 \\ 
		& & &  sc-SchNet & \textbf{0.08} $\pm$ 0.00 & \textbf{0.11} $\pm$ 0.01 \\
		&  \multirow{2}{*}{forces} & \multirow{2}{*}{kcal mol$^{-1}$ {\AA}$^{-1}$} & SchNet & 0.32 $\pm$ 0.01 & 0.46 $\pm$ 0.01 \\ 
		& & &  sc-SchNet & \textbf{0.31} $\pm$ 0.01 & \textbf{0.45} $\pm$ 0.01 \\ 
		\midrule
		
		\multirow{4}{*}{Ethanol (N=1k)} &  \multirow{2}{*}{energy} & \multirow{2}{*}{kcal mol$^{-1}$} & SchNet & 0.08 $\pm$ 0.00 & 0.12 $\pm$ 0.01\\ 
		& & &  sc-SchNet & \textbf{0.07} $\pm$ 0.00 & \textbf{0.11} $\pm$ 0.01 \\
		&  \multirow{2}{*}{forces} & \multirow{2}{*}{kcal mol$^{-1}$ {\AA}$^{-1}$} & SchNet & \textbf{0.29} $\pm$ 0.01 & \textbf{0.52}$\pm$ 0.01\\ 
		& & &  sc-SchNet & 0.30 $\pm$ 0.01 & 0.54 $\pm$ 0.01 \\ 
		\midrule
		\multirow{2}{*}{QM9 (N=110k)} & \multirow{2}{*}{$U_0$}  & \multirow{2}{*}{kcal mol$^{-1}$} & SchNet & \textbf{0.24} $\pm$ 0.01 & \textbf{0.43 }$\pm$ 0.01 \\
		& & &  sc-SchNet & \textbf{0.24} $\pm$ 0.01 & 0.46 $\pm$ 0.01 \\ 
		\bottomrule
	\end{tabular}
\end{table}

In the following, we demonstrate the performance of sc-SchNet for energy prediction as well as an analysis of how different datasets influence the representations.
We evaluate our model for energy prediction across chemical compound space using the QM9 benchmark~\cite{qm9_1,qm9_2} as well as for prediction of potential energy surfaces and force fields for the MD17 dataset~\cite{chmiela2018towards,Schnet_NIPS,schnet_JCP}.
Our models have been implemented using SchNetPack~\cite{Schnetpack}, which is based on the deep learning framework PyTorch~\cite{pytorch_paper}.

\subsection{Property Prediction}

We predict the internal energy $U_0$ for the QM9 benchmark which contains $\sim133k$ organic molecules composed of the chemical elements H, C, N, O and F.
Each experiment is repeated on three different data splits for both \emph{SchNet} and \emph{sc-SchNet}.
We use SGDR\cite{loshchilov_sgdr:_2016} (stochastic gradient descent with warm restarts) as learning rate scheduler with an episode length of 50 epochs.
We take 110k molecules from the entire dataset which we divide in 105k examples for training and 5k for validation.
The remaining data is left for testing.
We use a mini-batch size of 100 examples and 256 features for the atom-wise representations as well as six interaction blocks. The initial learning rate is $1\cdot10^{-4}$.

In case of the MD17 dataset, we study molecular dynamics trajectories of benzene, ethanol, salicylic acid and aspirin.
Here, the mini-batch size consists of 50 examples and the initial learning rate is $5\cdot10^{-4}$.
For every MD17 trajectory, we trained both networks on a combined loss of energies and forces, where the force model is obtained as the negative partial derivative of the energy mode with respect to the atomic positions (see \cite{Schnet_NIPS}). 

Table~\ref{tab:results} shows the mean absolute and root mean squared errors for predictions of energies and forces with the corresponding standard errors.
We observe that the performances of both models are comparable.
\textit{sc-SchNet} achieves slightly better performances on both MAE and RMSE for energy prediction on all studied molecules from MD17.
For molecular forces and predictions on QM9, the results agree with conventional \textit{SchNet} within the standard error.

\subsection{Interaction Analysis}

\begin{figure}[t]
	\centering
	\includegraphics[width=0.8\textwidth]{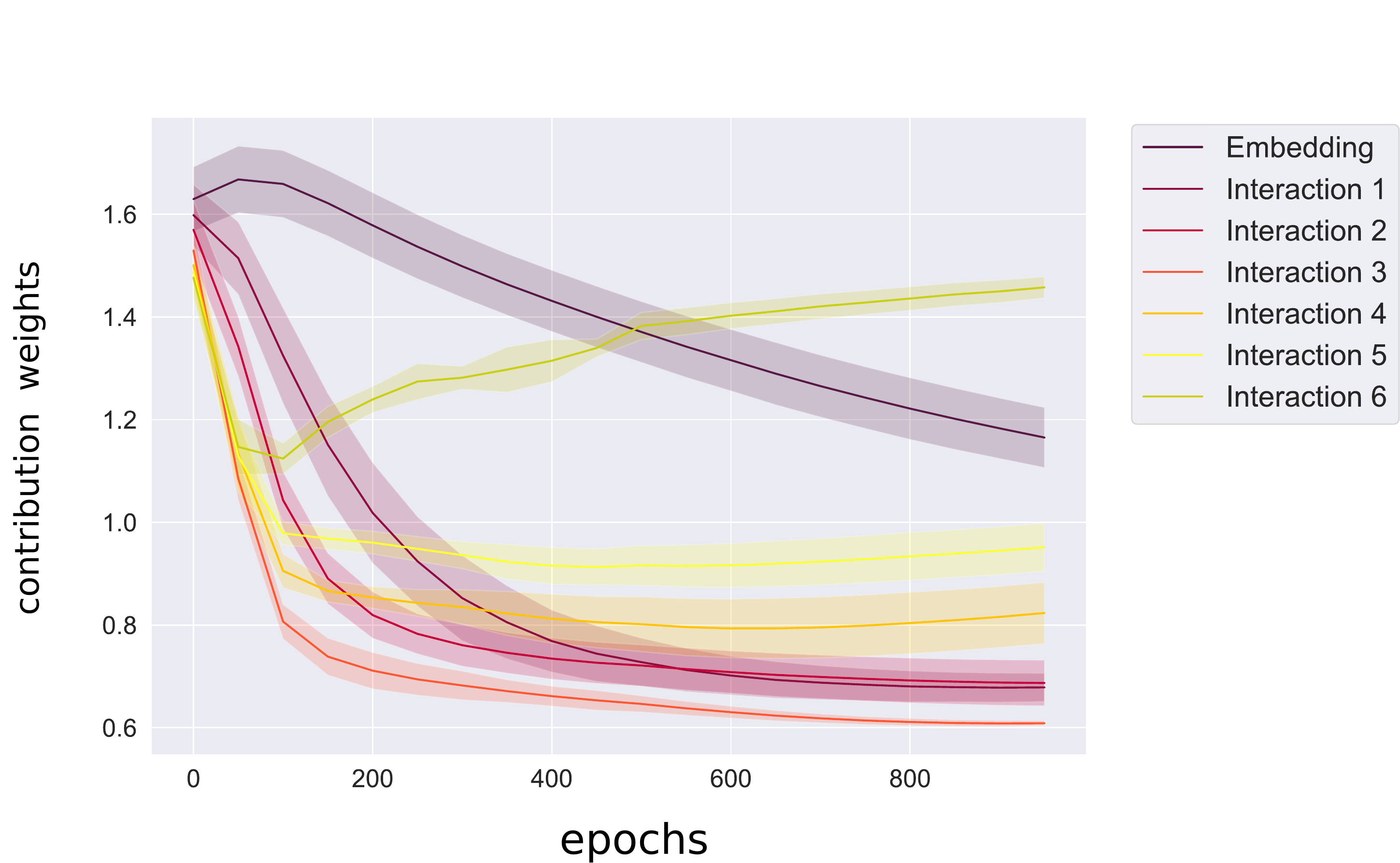}
	\caption[Evolution of contribution weights $w^{(l)}$ during Training]{Evolution of the contribution weights during training on the internal energy $U_0$ of the QM9 dataset.
	}
	\label{fig:figure_02}
\end{figure}

Having demonstrated that sc-SchNet achieves comparable prediction performance to SchNet, we go on to study the contribution weights according to Eq.~\ref{eq:weightsum_2} which can be interpreted as the importance of the information retrieved from the corresponding layer.
Fig.~\ref{fig:figure_02} shows their evolution during the training process for the QM9 dataset.
For each contribution weight, we report the average over three runs and the standard deviation as indicated by the shaded area surrounding each curve.

The most apparent aspects of Fig.~\ref{fig:figure_02} are the curves representing the initial embedding as well as the last interaction block, respectively, which contribute the most to the final representation.
While the initial embedding contains information about the chemical composition, which makes up for the majority of energy differences between molecules, the last interaction block contains the most complex representations of atomic environments.
Another important aspect is how the representation is obtained during training:
at first, most of the information is retrieved from the initial embedding and, to a lesser extent, the first and second interaction block.
However, the contribution of the early intermediates decreases steeply during the first epochs. 
This suggests a greater focus on the lower layers in the initial stages of training, where local, pair-wise features are useful for a rough fit of the energy.
For the network to propagate useful information to the higher layers, approximately 100 training epochs are required before the final interaction $\mathbf{v}_i^{(6)}$ exhibits rising contribution weights and surpasses the atom type embeddings after about 500 epochs.
While the intermediate interactions 1-5 are crucial to construct $\mathbf{v}_i^{(6)}$, they do not contribute that much to the final representation directly.

\begin{figure}[tb]
	\centering
	\includegraphics[width=\textwidth]{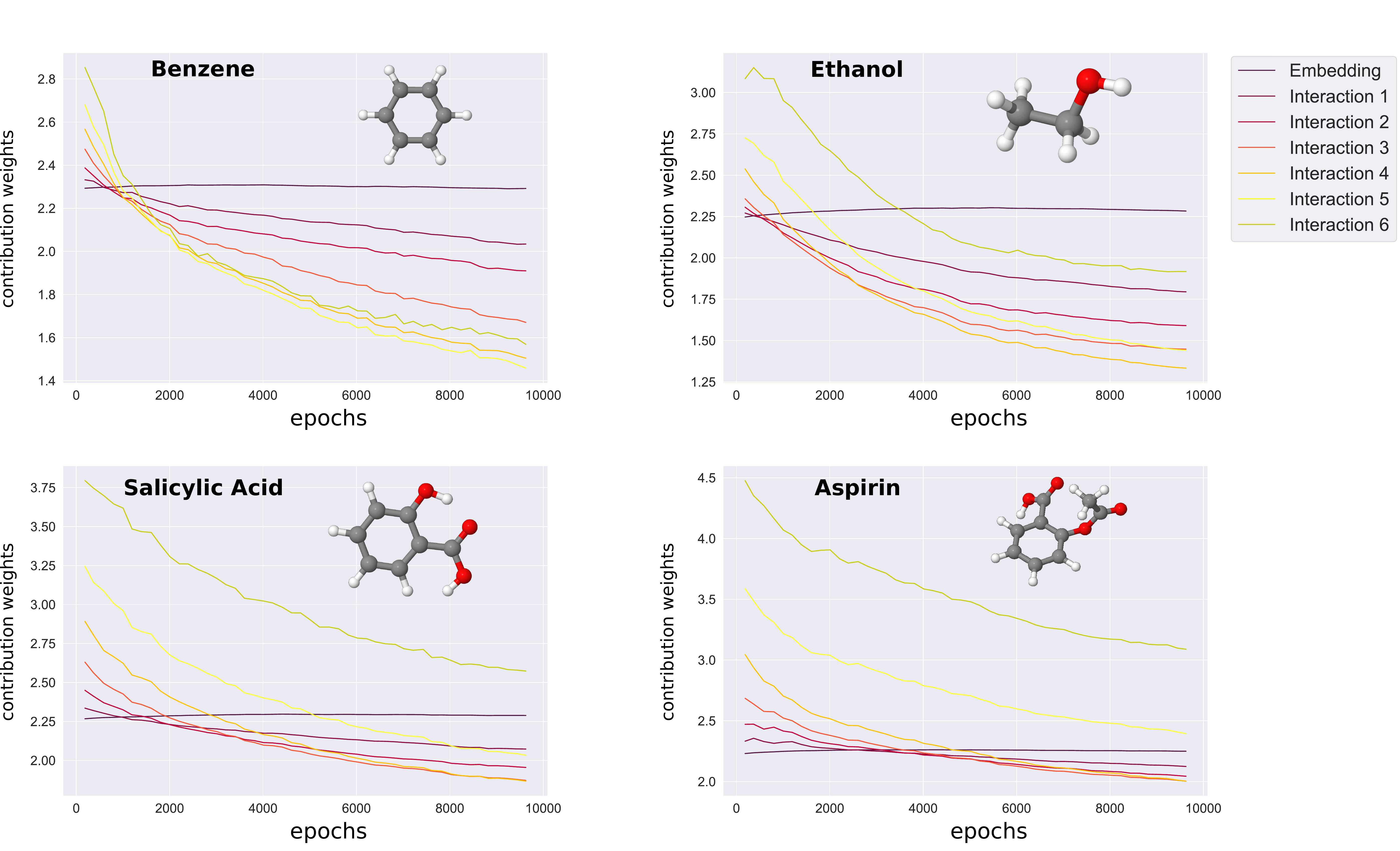}
	\caption[Activation weights of three different molecules]{Contribution weights for molecular dynamics simulations for four different molecules from the MD17 dataset: ethanol, benzene, salicylic acid and aspirin.} 
	\label{fig:figure_05}
\end{figure}

Fig.~\ref{fig:figure_05} presents the interaction contribution analysis for molecular dynamics trajectories from the MD17 dataset.
The contribution weights show a radically different behavior during the training processes compared to QM9.
In contrast to the QM9 results, the contribution of atom type embeddings remains approximately constant during training.
While QM9 contains a large variety of molecules, the chemical composition is the same for each MD17 trajectory, reducing the influence of the embedding on the energy.
However, the atom type still is important for the characterization of the interactions of an atom.
We observe this in particular for benzene which is quite stable due to its aromatic ring.
Since SchNet uses rotationally invariant, atom-wise features, it inherently accounts for the symmetries of the molecule.
Thus, the atom embedding for benzene corresponds directly to the position in the molecule, i.e. ring member or saturating hydrogen.

This becomes less apparent when moving to larger and more flexible molecules for which the interactions gain importance compared to the atom types.
The interaction features start off with relatively high contribution weights that decrease during training.
We conjecture that this is due to the fact that the relation between the geometry of the molecule and energy is first modeled by the interactions before sc-SchNet is able to resolve the interaction behavior of different atom types.
In contrast, the large energy changes in QM9 are due to changes in composition, so that the model can directly map this to atom type embeddings.

Comparing the curves of the contribution weights, in larger and more flexible molecules the higher interaction blocks are of greater importance since more complex chemical environments need to be represented.
For salicylic acid and aspirin, higher interactions contribute more than the atom type embeddings.
In other words, the complex interaction within the molecule cannot be resolved to less complex interactions as was the case for benzene and, partially, ethanol.
This demonstrates that the relative contribution of the atom embeddings and interactions during the learning process strongly depends on the molecular structure.

\section{Conclusions}

Atomistic end-to-end learning is not only able to yield fast and accurate predictions but can moreover be employed to obtain valuable insights from the learned representations~\cite{schnet_JCP,Book_chapter}.
In this work, we have introduced an extension to SchNet using skip layer connections in order to obtain a more transparent and interpretable model for latent representation of atomistic systems.
sc-SchNet not only obtains comparable predictions to SchNet but also allows us to study how the obtained atom-wise representations are assembled from atom type features as well as interaction corrections with different degree of complexity. 
We observe significant differences in the relative importance of the contribution weights depending on whether the model is trained across chemical compound space or on single trajectories.
Moreover, the interaction contributions reflect the size and flexibility of the molecule to be predicted.
In future studies, we will investigate whether the interaction coefficients encode information on the locality of different molecular properties. In addition, their relation to fundamental modes of molecular motion will be explored in the context of the MD17 molecules.

\section*{Acknowledgements}
This work was supported by the Federal Ministry of Education and Research (BMBF) for the Berlin Big Data
Center BBDC (01IS14013A) and the Berliner Zentrum für Maschinelles Lernen BZML (01IS18037A). MG acknowledges support provided by the European Union’s Horizon 2020 research and innovation program under the Marie Sk\l{}odowska-Curie grant agreement NO 792572.
Correspondence to KAN and KTS.

\bibliography{references}
\bibliographystyle{unsrt}
\end{document}